\documentclass[trackchanges,twocolumn]{aastex701}
\shorttitle{Interstellar Comet 3I/ATLAS}
\shortauthors{G. Ahuja \& S. Ganesh}
\submitjournal{RNAAS}

\begin{document}

\title{Effect of different Non-Gravitational accelerations on the trajectory of Interstellar Comet 3I/ATLAS}

\correspondingauthor{Goldy Ahuja}
\author[orcid=0009-0008-1809-3256,gname='Goldy',sname='Ahuja']{Goldy Ahuja}
\affiliation{Physical Research Laboratory, Ahmedabad, Gujarat-380009, India}
\affiliation{Indian Institute of Technology Gandhinagar, Palaj, Gujarat-382355, India}
\email[show]{goldy@prl.res.in, goldezz20@gmail.com}  

\author[orcid=0000-0002-7721-3827,gname=Shashikiran, sname=Ganesh]{Shashikiran Ganesh} 
\affiliation{Physical Research Laboratory, Ahmedabad, Gujarat-380009, India}
\email{shashi@prl.res.in}

\begin{abstract}

Comet C/2025 N1 or 3I/ATLAS is the third confirmed interstellar object. It has passed perihelion on 2025 October 29, and is currently on a path to leave the solar system. During its outbound journey, it will pass close to Jupiter at a distance of 0.358 au. NASA JPL \textsc{Horizons} has updated the non-gravitational parameters of the comet based on the CO$_2$ sublimation model, where $g(r)= 1/r^2$. In this research note, we use the non-gravitational accelerations from \textsc{Horizons} together with symmetric and asymmetric H$_2$O sublimation models derived using \texttt{Find\_Orb} software. We calculate the resulting perijove distances and compare them with our earlier results at epoch JD 2460867.5.  

\end{abstract}



\section{Introduction} 
Interstellar bodies are among the most sought-after objects for studying the possible history of other stellar systems. In July 2025, the third confirmed interstellar visitor, C/2025 N1 or 3I/ATLAS (hereafter 3I), was discovered at heliocentric and geocentric distances of 4.51 au and 3.50 au, respectively. The orbital eccentricity, $e$, is found to be around 6.1 and in a retrograde orbit having inclination, $i$, of around 175.1$^{\circ}$ \citep{3I_disc_1, 3I_disc_2}. It passed perihelion on 2025 October 29 at 1.357 au and is en route to a close encounter with Jupiter. 

In our earlier dynamical study of comet 3I \citep{Ahuja_2025}, we calculated the perijove distance using statistical clones at epoch JD 2460867.5 (2025 July 11, 00:00:00 TDB) before non-gravitational (hereafter NG) parameters were available. Assuming a water sublimation model (see \citet{Królikowska_2017}), we found that 87 clones out of 500 are entering the Hill radius of Jupiter for a representative NG acceleration of $1 \times 10^{-8}$ au day$^{-2}$ applied to both $A_1$ and $A_2$.

Recently, NASA JPL \textsc{Horizons}\footnote{\url{https://ssd.jpl.nasa.gov/tools/sbdb_lookup.html\#/?sstr=3I\%2FATLAS}} (hereafter \textsc{Horizons}) has updated the NG acceleration parameters at the epoch of JD 2460902.5 (2025-Aug-15 00:00:00 TDB), in which the CO$_2$ sublimation model is considered as the comet 3I has a CO$_2$-dominated coma \citep{Cordiner_2025,Lisse_2025}. Using these parameters, \textsc{Horizons} gives the predicted perijove distance as 0.35833-0.35834 au, which is just outside the Hill radius of Jupiter, 0.355 au. 

In this research note, we recompute the perijove distance of comet 3I using the updated NG accelerations from \textsc{Horizons}. We compare this distance with those calculated using symmetric and asymmetric H$_2$O sublimation models derived with \texttt{Find\_Orb} software. In Section \ref{method}, we outline the methodology for setting up the simulation, and in Section \ref{results}, we present the results and compare them with our earlier work \citep{Ahuja_2025}.

\section{Methodology}\label{method}
We adopt the same numerical setup and integration scheme as in our earlier work \citep{Ahuja_2025}, generating 500 statistical clones at the updated epoch JD 2460902.5 (2025 August 15, 00:00:00 TDB).
In addition to gravitational perturbations, comet 3I experiences NG accelerations due to outgassing. These accelerations are modelled following the standard Marsden formalism,
\begin{equation}
    F_i = A_i g(r) \quad i=1,2,3
\end{equation}
Here, $F_i$ are the radial, transverse, and normal components of the NG acceleration, $A_i$ are the corresponding components evaluated at 1 au from the Sun, and $g(r)$ is the semi-empirical sublimation function (see \citet{sosa_2011} and references therein).
\begin{equation}\label{gr}
    g(r) = \alpha \left(\frac{r}{r_0}\right)^{-m} \left[1 + \left(\frac{r}{r_0}\right)^n \right]^{-k}
\end{equation}
For CO$_2$ sublimation, the different parameters in equation (\ref{gr}) i.e., $\alpha = 1$, $m=2$, $n=0$, $k=0$, $r_0 = 1$ au, which gives $g(r) = 1/r^2$, and for H$_2$O sublimation, the values of constants are $\alpha = 0.1113$, $m=2.15$, $n=5.093$, $k=4.6142$, $r_0 = 2.808$ au (See \citet{Marsden_1973, sosa_2011, Królikowska_2017}). 

The NG accelerations for the different components are given by \textsc{Horizons} at the epoch of JD 2460902.5, based on the solution date of 08 Jan 2026, using a data arc from 2025 May 15 to 2025 Dec 26. The values are A$_1$ = (4.467 $\pm$ 0.128) $\times$ 10$^{-8}$ au day$^{-2}$, A$_2$ = (1.689 $\pm$ 0.205) $\times$ 10$^{-8}$ au day$^{-2}$ and A$_3$ = (-5.350 $\pm$ 0.352) $\times$ 10$^{-9}$ au day$^{-2}$. Similarly, using the \texttt{Find\_Orb} software \citep{Bill_gray_2022} and considering the data arc from 2025 May 08 to 2025 Dec 26, we have calculated various NG acceleration parameters, taking into account the H$_2$O sublimation model. For symmetric NG H$_2$O model, A$_1$ = (1.212 $\pm$ 0.144) $\times$ 10$^{-7}$ au day$^{-2}$, A$_2$ = (4.725 $\pm$ 0.346) $\times$ 10$^{-8}$ au day$^{-2}$ and A$_3$ = 0. For asymmetric NG H$_2$O model, A$_1$ = (-1.541 $\pm$ 0.178) $\times$ 10$^{-7}$ au day$^{-2}$, A$_2$ = (8.139 $\pm$ 0.359) $\times$ 10$^{-8}$ au day$^{-2}$ and A$_3$ = (-1.091 $\pm$ 0.045) $\times$ 10$^{-8}$ au day$^{-2}$. Here, we have not included the fourth NG parameter, DT (time offset at maximum brightness), in our simulation, as fits are poorly constrained by the available astrometry and lead to unstable NG solutions. 
        
\section{Results}\label{results}
\begin{figure*}
\plotone{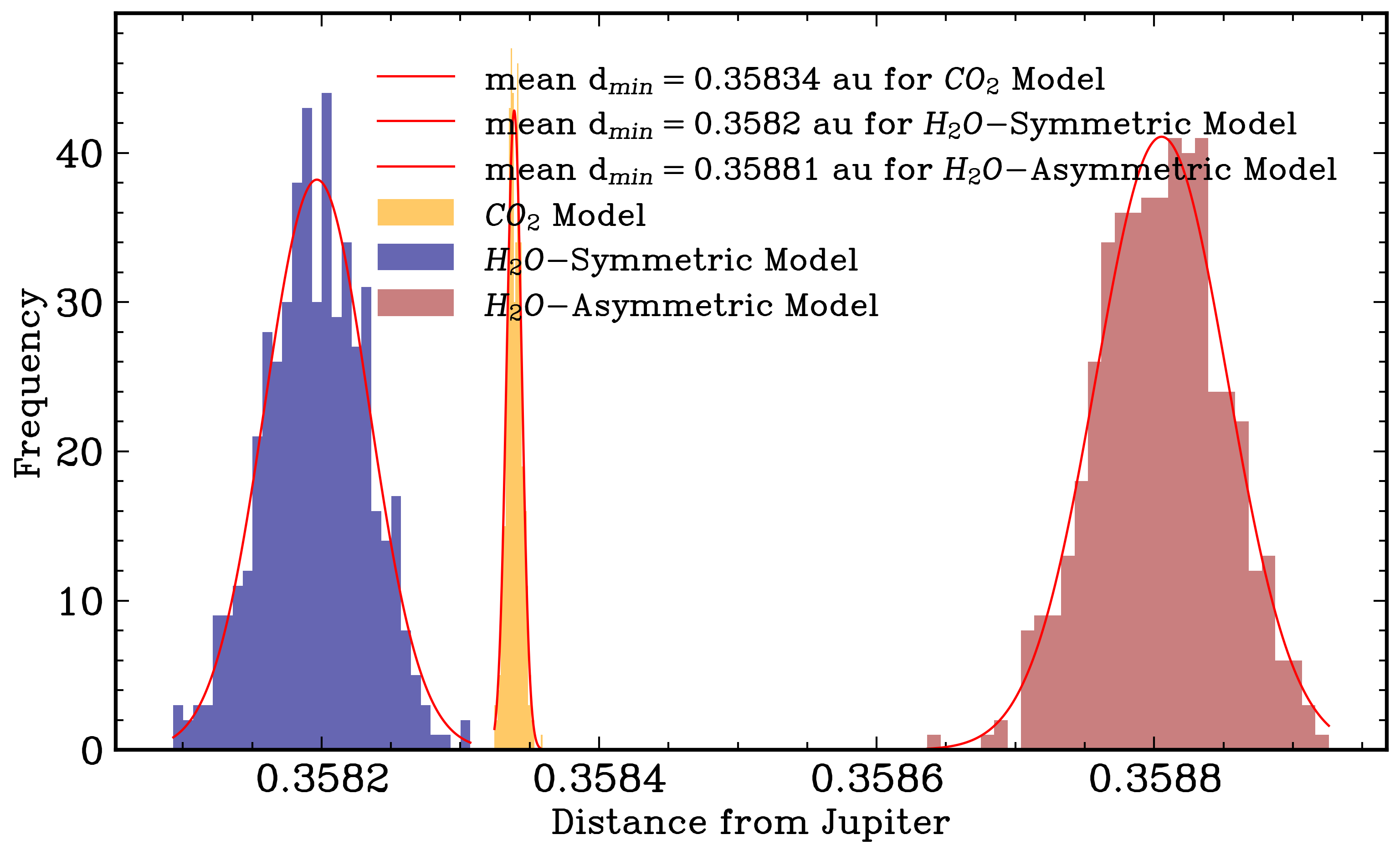}
\caption{Histogram of the perijove distances of 500 clones under different sublimation models. The Hill radius of Jupiter (not shown) is 0.355 au.}
\label{fig: histogram}
\end{figure*}

We find that for CO$_2$ sublimation, the histogram of the perijove distance lies between 0.35832-0.35835 au, which is similar to what \textsc{Horizons} reports. Using the two different H$_2$O sublimation models, the symmetric model gives the distances between 0.35810-0.35830 au, while for the asymmetric model, the distance is 0.35864-0.35892 au. The results shown in Figure \ref{fig: histogram} confirm that the most probable perijove distance is around 0.3582-0.3588 au, irrespective of the NG models adopted.

As we discussed in \citet{Ahuja_2025}, roughly 17 per cent (87 out of 500) of clones entered Jupiter's Hill radius using the orbital solution available at epoch JD 2460867.5. With improved astrometry and updated orbital parameters, the clone distribution shifts such that all clones have perijove distances beyond Jupiter's Hill radius.

\begin{acknowledgments}
Work at the PRL is supported by the Department of Space, Govt. of India. The computations were performed on the Param Vikram-1000 HPC Cluster of the PRL.
\end{acknowledgments}

\software{Astropy \citep{2013A&A...558A..33A,2018AJ....156..123A,2022ApJ...935..167A},  
          Astroquery \citep{astroquery_2019}, 
          Numpy \citep{Numpy_2020},
          Scipy \citep{2020SciPy-NMeth},
          Matplotlib \citep{Matplotlib_2007},
          Smplotlib \citep{smplotlib},
          Pandas \citep{pandas_2020},
          REBOUND \citep{rebound,reboundias15},
          Find\_Orb \citep{Bill_gray_2022}
          }




\bibliography{Reference}{}
\bibliographystyle{aasjournalv7}



\end{document}